\documentstyle[]{article}
\font\cero=cmss10 scaled 1728 \font\uno=cmssbx10 scaled 1200
\begin{document}
 \small{
\begin{flushleft}
{\cero Identically closed two-form for
covariant phase space
quantization of
Dirac-Nambu-Goto p-branes in a curved spacetime} \\[3em]
\end{flushleft}
{\sf R. Cartas-Fuentevilla}\\
{\it Instituto de F\'{\i}sica, Universidad Aut\'onoma de Puebla,
Apartado postal J-48 72570, Puebla Pue., M\'exico (rcartas@sirio.ifuap.buap.mx).}  \\[4em]

Using a fully covariant formalism given by Carter for the
deformation dynamics of p-branes governed by the Dirac-Nambu-Goto
action in a curved background, it is proved that the
corresponding Witten's phase space is endowed with a covariant
symplectic structure, which can serve as a starting point for a
phase space quantization of such objects. Some open
questions for further research are outlined. \\

\noindent Keywords: Symplectic structure, canonical quantization, p-branes.\\

\begin{center}
{\uno I. INTRODUCTION}
\end{center}
\vspace{1em}

A p-dimensional brane is a  relativistic extended object
propagating in a background spacetime manifold of some dimension
N. Such a dynamical system is defined in terms of fields with
support confined to a (p+1)-dimensional world sheet manifold that
the p-brane swept out in the course of its evolution, in such a
way that $N\geq (p+1)$. In this sense, a 0-brane is a point
particle, a 1-brane is a string, a 2-brane a domain wall or
membrane, and so on. The extreme case (N-1)-brane corresponds to
that of a continuous medium for which the confinement condition
is redundant. It is well known the very wide range of physically
diverse phenomena modeled as a p-brane of an appropriate
dimension. Originally, Dirac proposed his bubble-membrane as ``an
extensible model of the electron" \cite{1}, and Nambu and Goto
modeled hadronic matter using a relativistic string \cite{2}, and
later string theory was developed as a possible theory of strong
interactions \cite{3}. Subsequently, in the last years the theory
of extended objects is considered the point of departure in the
construction of the modern string/M-theory, which claims to
give a consistent unified description of all interactions
including gravity, and particularly a way of overcoming the
non-renormalizability of point-like theories.

In the last years a considerable amount of effort has been
devoted for developing a quantum field theory of extended objects
(which in fact, will constitute the ultimate framework for a
complete string/M-theory), however it has not yet been fully
developed. The quantization of extended objects is a very
complicated problem in physics because, among other things,
the theory is highly non-linear and the standard methods are not
directly applied, even for extended objects of simple topologies.
In this manner, the main purpose of the present work is to
explore the basic elements of one of such methods, the so called
phase space quantization that was originally introduced by Witten
\cite{4,5}, and that is based on a covariant description of the
standard canonical formalism. We shall restrict the present
analysis to the bosonic p-branes of the simplest topology, the
internally structureless p-dimensional generalization of the
Dirac-Nambu-Goto (DNG) membranes embedded in a curved spacetime,
for clarifying our basic ideas and to prepare the background for a
future treatment of physically more interesting cases.

In the next section, we summarize the DNG action for p-branes and the
covariant description of the corresponding deformations given by
Carter \cite{6}, which is essential for our present aims. In
Section III, using the concept of (self-)adjoint operators, we
derive a local continuity equation that will allow to identify a
bilinear form on deformations of classical solutions, which will
be connected directly to the wanted canonical structure on the
phase space; the later will be explained in Section IV. In
Section V the closeness of the bilinear form previously
constructed is proved. We finish in Section VI with some
discussions about our results and future extensions. \\[2em]

\noindent {\uno II. THE DNG ACTION AND FIRST ORDER DEFORMATIONS}
\vspace{1em}

The DNG action for p-branes in an arbitrary curved background is
given by an integral proportional to the area swept out by the
world sheet:
\begin{equation}
     {\cal S} = \int \widetilde{L} d \widetilde{\Sigma}, \quad \widetilde{L} =
- m^{\rho +1},
\end{equation}
where $m$ is a fixed parameter, and $d \widetilde{\Sigma}$ is the
surface measure element induced on the world sheet by the
background metric $(g_{\mu\nu})$. Using the decomposition
$g^{\mu\nu} = \eta^{\mu\nu} + \bot^{\mu\nu}$ of the background
metric in terms of the tangential and orthogonal projectors to the
world sheet, with the properties
\begin{equation}
     \eta^{\mu} {_{\rho}} \eta^{\rho} {_{\nu}} = \eta^{\mu}
{_{\nu}}, \quad \bot^{\mu} {_{\rho}} \bot^{\rho} {_{\nu}} =
\bot^{\mu} {_{\nu}}, \quad \eta^{\mu} {_{\rho}} \bot^{\rho}
{_{\nu}} = 0,
\end{equation}
the dynamical equations of motion of the action (1) are
expressible as the harmonicity condition
\begin{equation}
     \widetilde{\nabla}_{\mu} \eta^{\mu} {_{\nu}} = 0,
\end{equation}
where $\widetilde{\nabla}_{\mu} \equiv \eta^{\nu} {_{\mu}}
\nabla_{\nu}$ is the tangential covariant differentiation
($\nabla_{\nu}$ denotes Riemannian covariant differentiation with
respect to the background metric), which is the only meaningful
for tensor fields whose support is confined to the world sheet.

According to the Carter approach \cite{6}, the linearized dynamics
of first order deformations of the p-branes under consideration,
can be described in terms of the infinitesimal displacement
vector field $\xi^{\mu}$ by the equations:
\begin{equation}
     (\bot^{\mu} {_{\lambda}} \widetilde{\nabla}_{\nu}
\widetilde{\nabla}^{\nu} - 2 K_{\lambda} {^{\nu\mu}}
\widetilde{\nabla}_{\nu} + \bot^{\mu} {_{\nu}} \eta^{\rho\sigma}
{\cal R}_{\rho} {^{\nu}}_{\sigma\lambda}) \xi^{\lambda} \equiv
({\cal P} \xi^{\lambda})^{\mu} = 0,
\end{equation}
where $K_{\mu\nu} {^{\rho}} = \eta^{\lambda}{_{\nu}}
\widetilde{\nabla}_{\mu} \eta^{\rho}{_{\lambda}}$ is the {\it
second fundamental tensor}, and ${\cal R}_{\mu\rho\nu\sigma}$
denotes the Riemann curvature of the background spacetime. Note
that the linear operator $\cal P$ is taking vector fields into
themselves. $\xi^{\mu}$ can be identified with the corresponding
first order change in the coordinate fields $x^{\mu}$ of the
background, $\xi^{\mu} = \delta_{L} x^{\mu}$, where $\delta_{L}$
denotes a {\it Lagrangian} deformation in the sense of being
defined with respect to a reference system comoving with the
relevant displacement \cite{6}. For fields with support confined
to the world sheet, their Lagragian deformation reduces to the
Lie differentiation with respect to the displacement vector
$\xi^{\mu} ({\cal L}_{\xi})$. For fields whose support is not
confined to the world sheet, there will be a {\it fixed point}
differential $(\delta_{E})$ in such a way that $\delta_{L} =
{\cal L}_{\xi} + \delta_{E}$. In particular, in the Carter
approach is considered that {\it the background metric is fixed},
$\delta_{E} g_{\mu\nu} = 0$, and then
\begin{equation}
     \delta_{L} g_{\mu\nu} = {\cal L}_{\xi} g_{\mu\nu} =
\nabla_{\mu} \xi_{\nu} + \nabla_{\nu} \xi_{\mu}, \quad \delta_{L}
g^{\mu\nu} = - (\nabla^{\mu} \xi^{\nu} + \nabla^{\nu} \xi^{\mu}),
\end{equation}
which will be useful below. For more details about this section,
see directly Ref.\ \cite{6} (and also \cite{7}, particularly the
sections 4.1--4.4). \\[2em]

\noindent {\uno III. ADJOINT OPERATORS AND COVARIANTLY CONSERVED
CURRENTS}
\vspace{1em}

In this Section we shall construct from Eq.\ (4), and using the
concept of self-adjoint operators, a bilinear form on the
displacement vector $\xi^{\mu}$ and their derivatives, which will
be world sheet covariantly conserved. \\[2em]

\noindent {\uno 3.1. Self-adjoint operators and local continuity
laws}
\vspace{0.5cm}

The general relationship between adjoint operators and covariantly
conserved currents has been already given in previous works (see
for example \cite{8} and references cited therein), however we
shall discuss it in this section for completeness.

If ${\cal P}$ is a linear partial differential operator which
takes matrix-valued tensor fields into themselves, then, the
adjoint operator of ${\cal P}$, is that operator ${\cal
P}^{\dag}$, such that
\begin{equation}
      f^{\rho\sigma ...}[{\cal P}(g_{\mu\nu ...})]_{\rho\sigma ...}
- [{\cal P}^{\dag}(f^{\rho\sigma ...})]^{\mu\nu ...}g_{\mu\nu ...}
 = \nabla_{\mu} {\cal J}^{\mu},
\end{equation}
where ${\cal J}^{\mu}$ is some vector field. From this
definition, if $\cal Q$ and $\cal R$ are any two linear
operators, one easily finds the following properties:
\[
   ({\cal Q}{\cal R})^{\dag} = {\cal R}^{\dag}{\cal Q}^{\dag}, \qquad
({\cal Q} + {\cal R})^{\dag} = {\cal Q}^{\dag} + {\cal R}^{\dag},
\]
and in the case of a function $F$,
\[
    F^{\dag} = F,
\]
which will be used implicitly below.

From Eq.\ (6) we can see that this definition automatically
guarantees that, if ${\cal P}$ is a self-adjoint operator (${\cal
P}^{\dag} = {\cal P}$), and the fields $f$  and $g$ correspond to
a pair of solutions admitted by the linear system ${\cal P} (f) =
0 = {\cal P}(g)$, then we obtain the continuity law $\nabla_{\mu}
{\cal J}^{\mu}=0$, which establishes that ${\cal J}^{\mu}$ is a
covariantly conserved current, bilinear on the fields $f$ and $g$.
This fact means that for any self-adjoint homogeneous equation
system, one can always to construct a conserved current. Although
this result has been established assuming only tensor fields and
the presence of a single equation, such a result can be extended
in a direct way to equations involving spinor fields, matrix
fields, and the presence of more that one field. \\[2em]

\noindent {\uno 3.2. Self-adjointness of the linearized dynamics}
\vspace{0.5cm}

In this section we shall demonstrate that the operator $\cal P$
in Eq.\ (4) is indeed self-adjoint. With this purpose, let
$\acute{\xi}_{\mu}$ and $\xi_{\lambda}$ be two vector fields
(which will be identified as a pair of solutions of Eq.\ (4)), and
following the definition (6), we need to analyze the scalar
expression $\acute{\xi}_{\mu} ({\cal P}\xi^{\lambda})^{\mu}$, and
thus the following identities will be useful (and very easy to
verify):
\begin{eqnarray}
     \acute{\xi}_{\mu} \bot^{\mu}_{\lambda}
\widetilde{\nabla}_{\nu} \widetilde{\nabla}^{\nu} \xi^{\lambda}
\!\! & \equiv & \!\! \widetilde{\nabla}_{\nu} {\cal J}^{\nu}_{1}
+ \{ [ \bot^{\mu}_{\lambda} \widetilde{\nabla}_{\nu}
\widetilde{\nabla}^{\nu} + 2 (\widetilde{\nabla}^{\nu}
\bot^{\mu}_{\lambda}) \widetilde{\nabla}_{\nu} +
(\widetilde{\nabla}_{\nu} \widetilde{\nabla}^{\nu}
\bot^{\mu}_{\lambda})] \acute{\xi}_{\mu} \} \xi^{\lambda}, \\
     - 2 \acute{\xi}_{\mu} K_{\lambda} {^{\nu\mu}}
\widetilde{\nabla}_{\nu} \xi^{\lambda} \!\! & \equiv & \!\!
\widetilde{\nabla}_{\nu} {\cal J}^{\nu}_{2} + 2 \{ [ K_{\lambda}
{^{\nu\mu}} \widetilde{\nabla}_{\nu} + (\widetilde{\nabla}_{\nu}
K_{\lambda} {^{\nu\mu}})] \acute{\xi}_{\mu} \} \xi^{\lambda},
\end{eqnarray}
where ${\cal J}^{\nu}_{1} = \bot^{\mu}_{\lambda} \acute{\xi}_{\mu}
\widetilde{\nabla}^{\nu} \xi^{\lambda} - \widetilde{\nabla}^{\nu}
(\bot^{\mu}_{\lambda} \acute{\xi}_{\mu}) \xi^{\lambda}$, and
${\cal J}^{\nu}_{2} = - 2 K_{\lambda} {^{\nu\mu}}
\acute{\xi}_{\mu} \xi^{\lambda}$. The first term
$\bot^{\nu}_{\lambda} \widetilde{\nabla}_{\nu}
\widetilde{\nabla}^{\nu} \acute{\xi}_{\mu}$ in the $\{ \}$ terms
in Eq.\ (7) is directly the first term of $({\cal P}
\acute{\xi}_{\mu})_{\lambda}$. We add (and subtract) the
complementary terms $(- 2 K^{\mu\nu\lambda}
\widetilde{\nabla}_{\nu} + \bot^{\lambda}_{\nu} \eta^{\rho\sigma}
{\cal R}_{\rho} {^{\nu}} {_{\sigma}} {^{\mu}}) \acute{\xi}_{\mu}$
for making $({\cal P} \xi^{'}_{\mu})_{\lambda}$ on the right
hand-side of Eq.\ (7) (following Eq.\ (4)), and to obtain, after
some arrangements, that
\begin{eqnarray}
     \acute{\xi}_{\mu} ({\cal P} \xi^{\lambda})^{\mu} \!\! & = & \!\!
({\cal P} \acute{\xi}_{\mu})_{\lambda} \xi^{\lambda} +
\widetilde{\nabla}_{\nu} ({\cal J}^{\nu}_{1} +  {\cal
J}^{\nu}_{2}) + 2 [K^{\mu\nu} {_{\lambda}} + K_{\lambda}
{^{\nu\mu}} + \widetilde{\nabla}^{\nu} \bot^{\mu}_{\lambda}]
(\widetilde{\nabla}_{\nu} \acute{\xi}_{\mu}) \xi^{\lambda}
\nonumber \\
     \!\! & & \!\! + (\widetilde{\nabla}_{\nu} \widetilde{\nabla}^{\nu}
\bot^{\mu}_{\lambda} + 2 \widetilde{\nabla}_{\nu} K_{\lambda}
{^{\nu\mu}}) \acute{\xi}_{\mu} \xi^{\lambda} + \eta^{\rho\sigma}
(\bot^{\mu}_{\nu} {\cal R}_{\rho} {^{\nu}} {_{\sigma\lambda}} -
\bot_{\lambda\nu} {\cal R}_{\rho} {^{\nu}} {_{\sigma}} {^{\mu}})
\acute{\xi}_{\mu} \xi^{\lambda},
\end{eqnarray}
where it remains to demonstrate that, in according to Eq.\ (6), if
$\cal P$ is self-adjoint, then the last three terms in Eq.\ (9)
correspond at most to a pure divergence, such as the second term.
This is effectively the case; using the definition of $K^{\mu\nu}
{_{\lambda}}$ and the first of properties (2), it is very easy to
find that $K^{\mu\nu} {_{\lambda}} + K_{\lambda} {^{\nu\mu}} +
\widetilde{\nabla}^{\nu} \bot^{\mu}_{\lambda} =
\widetilde{\nabla}^{\nu} \eta^{\mu}_{\lambda} +
\widetilde{\nabla}^{\nu} \bot^{\mu}_{\lambda} = 0$, and thus the
third term in Eq.\ (9) vanishes. Furthermore, using the definition
$2 \nabla_{[\mu} \nabla_{\nu ]} {\cal A}_{\rho} = {\cal
R}_{\mu\nu\rho} {^{\lambda}} {\cal A}_{\lambda}$ and the relation
$\bot^{\mu}_{\nu} = g^{\mu}_{\nu} -\eta^{\mu}_{\nu}$, one can
demonstrate easily that
\begin{equation}
     \eta^{\sigma\rho} (\bot^{\mu}_{\nu} {\cal R}_{\rho} {^{\nu}}
{_{\sigma\lambda}} - \bot_{\lambda\nu} {\cal R}_{\rho} {^{\nu}}
{_{\sigma}} {^{\mu}}) \acute{\xi}_{\mu} \xi^{\lambda} =
\widetilde{\nabla}_{\nu} {\cal J}^{\nu}_{3} +
(\widetilde{\nabla}^{\nu} \bot^{\mu}_{\lambda} + 2 K_{\lambda}
{^{\nu\mu}}) \widetilde{\nabla}_{\nu} (\acute{\xi}_{\mu}
\xi^{\lambda}),
\end{equation}
where ${\cal J}^{\nu}_{3} = \widetilde{\nabla}_{\mu}
(\acute{\xi}^{\nu} \xi^{\mu} - \acute{\xi}^{\mu} \xi^{\nu})$.
Hence, from Eqs.\ (9) and (10), we have finally
\begin{equation}
     \acute{\xi}_{\mu} ({\cal P} \xi^{\lambda})^{\mu} = ({\cal P}
\acute{\xi}_{\mu})_{\lambda} \xi^{\lambda} +
\widetilde{\nabla}_{\nu} \widetilde{\cal J}^{\nu},
\end{equation}
where $\widetilde{\cal J}^{\nu} = {\cal J}^{\nu}_{1} + {\cal
J}^{\nu}_{2} + {\cal J}^{\nu}_{3} + (\widetilde{\nabla}^{\nu}
\bot^{\mu}_{\lambda} + 2 K_{\lambda} {^{\nu\mu}})
\acute{\xi}_{\mu} \xi^{\lambda}$, which has a remarkable
simplification by substituting  the explicit forms of ${\cal
J}_{1}$, ${\cal J}_{2}$, and ${\cal J}_{3}$:
\begin{equation}
     \widetilde{\cal J}^{\nu} = (\eta^{\mu\nu} \bot_{\sigma\rho} +
2 \eta^{\nu} {_{[\sigma}} \eta_{\rho ]} {^{\mu}})
[\acute{\xi}^{\sigma} \widetilde{\nabla}_{\mu} \xi^{\rho} -
(\widetilde{\nabla}_{\mu} \acute{\xi}^{\rho}) \xi^{\sigma}].
\end{equation}
Therefore, Eq.\ (11) has the form of (6) with ${\cal P}^{\dag} =
{\cal P}$ and $\widetilde{\cal J}^{\nu}$ will be world sheet
covariantly conserved
\begin{equation}
     \widetilde{\nabla}_{\nu} \widetilde{\cal J}^{\nu} = 0,
\end{equation}
if $\acute{\xi}_{\mu}$ and $\xi_{\lambda}$ correspond to a pair
of solutions of the linearized dynamics: $({\cal P}
\xi^{\lambda})_{\mu} = 0 = ({\cal P}
\acute{\xi}_{\mu})_{\lambda}$. The current $\widetilde{\cal
J}^{\nu}$ obtained in (12) is essentially that obtained by Carter
\cite{6} directly from the second order variation theory and it is
understood as a conventional symplectic Noetherian current on the
world sheet. Hence, we have up to here only reproduced by another
way, part of the results obtained in \cite{6}. However, the main
contribution of the present work is contained from the next
section, where we shall give, following Witten \cite{4,5}, a
nonconventional physical meaning to $ \widetilde{\cal J}^{\nu}$
on the covariant phase space, which will lead finally to the
construction of a symplectic structure for the theory. For
simplifying our calculations, we shall work in the orthogonal
gauge
\begin{equation}
     \eta^{\mu} {_{\nu}} \xi^{\nu} = 0,
\end{equation}
which removes the nonphysically observable tangential projection
of the deformation; in this gauge $ \widetilde{\cal J}^{\nu}$ has
an additional simplification \cite{6}:
\begin{equation}
      \widetilde{\cal J}^{\nu} = \acute{\xi}^{\rho}
      \widetilde{\nabla}^{\nu} \xi_{\rho} - (\widetilde{\nabla}^{\nu}
      \acute{\xi}^{\rho}) \xi_{\rho}.
\end{equation}
\\[2em]

\noindent {\uno IV. WITTEN PHASE SPACE}
\vspace{1em}

In according to Witten \cite{4,5}, in a given physical theory,
{\it the classical phase space is the space of solutions of the
classical equations of motion}, which corresponds to a manifestly
covariant definition. Based on this definition, the idea of
giving a covariant description of the canonical formalism
consists in describing Poisson brackets of the theory in terms of
a symplectic structure on such a phase space in a covariant way,
instead of choosing p's and q's. Strictly speaking, a symplectic
structure is a (non degenerate) closed two-form on the phase
space; hence, for working in this scheme an exterior calculus
associated with the phase space is fundamental. We summarize and
adjust  all these basic ideas about phase space quantization
given in Ref.\ \cite{4,5} for the case of p-branes treated here.

First, the phase space of DNG p-branes is the space of solutions
of Eqs.\ (3), and we shall call it $Z$. Any (unperturbed)
background quantity such as the background and internal metrics,
the projection tensors, etc., will be associated with zero-forms
on $Z$ \cite{4,5}. The Lagrangian deformation $\delta_{L}$ acts
as an exterior derivative on $Z$, taking $k$-forms into
$(k+1)$-forms, and it should satisfy
\begin{equation}
     \delta^{2}_{L} = 0,
\end{equation}
and the Leibniz rule
\[
     \delta_{L} (A B) = \delta_{L} A B + (-1)^{A} A\delta_{L} B.
\]
In particular, $\xi^{\mu} = \delta_{L} x^{\mu}$ is the exterior
derivative of the zero-form $x^{\mu}$, and corresponds to an
one-form on $Z$, and thus is an anticommutating object:
$\xi^{\mu} \xi^{\lambda} = - \xi^{\lambda} \xi^{\mu}$. In
according to (16), $\xi^{\mu}$ will be closed, $\delta_{L}
\xi^{\mu} = \delta^{2}_{L} x^{\mu} = 0$, which is evident from the
explicit form of $\delta_{L} \xi^{\mu}$ given in \cite{6}:
\begin{equation}
     \delta_{L} \xi^{\mu} = - \Gamma^{\mu}_{\lambda\nu} \xi^{\lambda}
\xi^{\nu} = 0,
\end{equation}
which vanishes because of the symmetry of the background
connection $\Gamma^{\mu}_{\lambda\nu}$ in its indices $\lambda$
and $\nu$ and the anticommutativity of the $\xi^{\lambda}$ on
$Z$.

In this manner, $\widetilde{\cal J}^{\nu}$ in Eq.\ (15), being
bilinear in the one-forms $\acute{\xi}^{\rho}$ and $\xi_{\rho}$,
corresponds to a two-form on $Z$; since $\widetilde{\nabla}_{\mu}$
depends only an unperturbed background and world sheet quantities
(zero-forms), $\widetilde{\nabla}_{\mu} \xi_{\rho}$ will be an
one-form on $Z$, such as $\xi_{\rho}$. The bilinear product
$\acute{\xi}^{\rho} \widetilde{\nabla}_{\mu} \xi_{\rho}$ must be
understood strictly as a wedge product of one-forms on $Z$:
$\acute{\xi}^{\rho} \wedge \widetilde{\nabla}_{\mu} \xi_{\rho}$;
however, we avoid the explicit use of $\wedge$ (such as Ref.\
\cite{4,5}), taking into account always that for differential
forms $A B = (-1)^{A B} BA$.

With these ideas, we can set $\acute{\xi}^{\rho} = \xi^{\rho}$ in
Eq.\ (15), without loosing generality, and considering that
$(\widetilde{\nabla}_{\mu} \xi^{\rho}) \xi_{\rho} = - \xi_{\rho}
\widetilde{\nabla}_{\mu} \xi^{\rho}$, one obtains essentially the
following two-form on $Z$:
\begin{equation}
     \widetilde{\cal J}^{\nu} = \xi^{\rho} \widetilde{\nabla}^{\nu}
     \xi_{\rho}.
\end{equation}
\\[2em]

\noindent {\uno V. THE SYMPLECTIC STRUCTURE ON Z}
\vspace{1em}

As mentioned, a symplectic structure is a closed two-form on $Z$.
The closeness means that the exterior derivative of such two-form
vanishes on $Z$. In this section we shall construct, from the
two-form (18), a symplectic structure for DNG p-branes.

Since the world sheet is an orientable manifold, integration is
defined in straightforward way, and we can construct, following
Witten \cite{4,5}, the following two-form on $Z$:
\begin{equation}
     \omega \equiv \int_{\Sigma} \sqrt{- \gamma} \widetilde{\cal
J}^{\mu} d \widetilde{\Sigma}_{\mu},
\end{equation}
where $\gamma$ is the determinant of the world sheet metric and
$\Sigma$ is a spacelike section of the world sheet manifold and
corresponds to an initial value p-surface for the configuration
of the p-brane; $d \widetilde{\Sigma}_{\mu}$ is the surface
measure element of $\Sigma$, and is normal on $\Sigma$ and
tangent to the world sheet. Employing Green's theorem in an usual
way (see for example Eq.\ (95) in \cite{7}), the world sheet
current conservation law (13) ensures that $\omega$ in (19) is
independent on the choice of $\Sigma$:
\begin{equation}
     \int_{\Sigma} \sqrt{- \gamma} \widetilde{\cal
J}^{\mu} d \widetilde{\Sigma}_{\mu} = \int_{\Sigma^{'}} \sqrt{-
\gamma} \widetilde{\cal J}^{\mu} d \widetilde{\Sigma}_{\mu}^{'},
\end{equation}
and is, in particular, Poincar\'{e} invariant. We shall
demonstrate now that $\omega$ is indeed a closed two-form on $Z$.
From Eq.\ (19), the exterior derivative of $\omega$ is given by
\begin{equation}
     \delta_{L} \omega = \int_{\Sigma} [ (\delta_{L} \sqrt{- \gamma})
\widetilde{\cal J}^{\mu} + \sqrt{- \gamma} \delta_{L}
\widetilde{\cal J}^{\mu} ] d \widetilde{\Sigma}_{\mu};
\end{equation}
considering that $\delta_{L} \sqrt{- \gamma} = 0$ corresponds to
the first order action variation (see Eqs.\ (1) and (3)), the
closeness of $\omega$ holds if $\widetilde{\cal J}^{\mu}$ itself
is closed. For demonstrating this property, we rewrite the
current (18) as $\widetilde{\cal J}^{\mu} = \eta^{\mu\nu} {\cal
J}_{\nu}$, where ${\cal J}_{\nu} \equiv \xi_{\rho} \nabla_{\nu}
\xi^{\rho}$, and let us show first that ${\cal J}_{\nu}$ is an
{\it exact} two-form on $Z$.

The Lagrangian variation of tensor fields in the Carter approach
(and that corresponds to the exterior derivative of such fields
on $Z$ in the present approach), is defined in terms of the
affinely parameterised geodesic equation
\begin{equation}
     \xi^{\nu} \nabla_{\nu} \xi^{\mu} = 0,
\end{equation}
which is fully equivalent to Eq.\ (134) in \cite{7}; $\xi^{\mu} =
\delta_{L} x^{\mu}$ is the affinely normalized tangent vector
along the geodesic. In particular, the exterior derivative of the
one-form $\xi_{\mu} \equiv g_{\mu\nu} \xi^{\nu}$ is given by
\begin{equation}
     \delta_{L} \xi_{\mu} = (\delta_{L} g_{\mu\nu}) \xi^{\nu} +
g_{\mu\nu} \delta_{L} \xi^{\nu} = (\nabla_{\mu} \xi_{\nu} +
\nabla_{\nu} \xi_{\mu}) \xi^{\nu} = (\nabla_{\mu} \xi_{\nu})
\xi^{\nu} = - {\cal J}_{\mu},
\end{equation}
where Eqs.\ (5), (17), and (22) have been used. Hence, Eq.\ (23)
shows that ${\cal J}_{\mu}$ is an {\it exact} two-form
(corresponds to the exterior derivative of an one-form), and
automatically a closed two-form, in view of (16):
\begin{equation}
     \delta_{L} {\cal J}_{\mu} = - \delta_{L}^{2} \xi_{\mu} = 0.
\end{equation}
In this manner $\widetilde{\cal J}^{\mu} = - \eta^{\mu\nu}
\delta_{L} \xi_{\nu}$ on $Z$, and
\begin{equation}
     \delta_{L} \widetilde{\cal J}^{\mu} = - (\delta_{L}
\eta^{\mu\nu}) \delta_{L} \xi_{\nu},
\end{equation}
in virtue of (24). However, from the property (16) and the
Leibniz rule, Eq.\ (25) can be rewritten as
\begin{equation}
     \delta_{L} \widetilde{\cal J}^{\mu} = \delta_{L} [(\delta_{L}
\eta^{\mu\nu}) \xi_{\nu}],
\end{equation}
and considering that $\delta_{L} \eta^{\mu\nu} = - \eta^{\mu\rho}
\eta^{\nu\sigma} \delta_{L}g_{\rho\sigma}$ \cite{6,7},
\begin{equation}
     (\delta_{L}\eta^{\mu\nu}) \xi_{\nu} = - \eta^{\mu\rho} \delta
g_{\rho\sigma} (\eta ^{\nu\sigma} \xi_{\nu}) = 0,
\end{equation}
in according to the orthogonality condition (14). Thus, Eqs.\ (26)
and (27) imply that $\delta_{L} \widetilde{\cal J}^{\mu} = 0$, and
$\omega$ is identically closed on $Z$, in according to Eq.\ (21).
Therefore, (19) is our wanted symplectic structure on $Z$ for DNG
p-branes.

It remains to discuss the gauge invariance properties of
$\omega$. Since all fields appearing in the definitions of
$\omega$ transform homogeneously like tensors (which is precisely
a virtue of the Carter formalism for deformations employed here),
$\omega$ is invariant under spacetime diffeomorphisms. Similarly,
$\omega$ involves integration of a world sheet scalar density,
and then is invariant under world sheet reparametrizations. In
this manner, our symplectic structure $\omega$ {\it inherits} the
covariant properties of the deformation formalism, from which it
has been constructed.  \\[2em]

\noindent {\uno VII. REMARKS AND PROSPECTS}
\vspace{1em}

In this manner, we have obtained a symplectic structure for DGN
p-branes propagating in an arbitrary curved background. Such a
structure is covariant in the strong sense of being expressed in
terms only of ordinary tensors, and not of internal world sheet
degrees of freedom. It is important to emphasize that, in
accordance with the above results, $\omega$ emerges in a natural
way and without any restrictions or additional assumptions, which
suggests a deeper research of the symplectic structure $\omega$,
with the idea that such a Hamiltonian scheme serves as the base of
a covariant canonical quantization of such objects on a curved
background. Specifically the issue of the degenerate directions and 
the existence of  {\it global} symplectic potentials on the phase space 
of the theory 
has been considered in a recent paper \cite{9}, and additionally
a {\it weakly} covariant description of the phase space for a restricted class
of topological defects has been also given in \cite{10}.

It is opportune to mention that the restriction on the background metric
considered in the present letter (see paragraph before Eq.(5)), has been
dropped in a recent work by R. Battye and B. Carter \cite{11}, and works
along these lines are in progress for overcoming such a
limitation. Finally, physically more interesting 
cases will be treated elsewhere following the ideas presented here. \\[2em]

\begin{center}
{\uno ACKNOWLEDGMENTS}
\end{center}
\vspace{1em}

This work was supported by CONACyT and the Sistema Nacional de
Investigadores (M\'exico). The author wants to thank H. Garcia
Compean for drawing my interest to the study of branes, and 
G. F. Torres del Castillo for discussions. The author also thanks
the referee for drawing my attention on Ref.\cite{11}.
\\[2em]

}
\end{document}